\documentclass[aps,preprint,superscriptaddress,showpacs]{revtex4}
\usepackage{epsfig}
\newcommand{\beq}{\begin{equation}} 
\newcommand{\eeq}{\end{equation}} 
\newcommand{\fr}{\frac}

\begin{document}            

\title{Application of the Trace Formula in Pseudointegrable Systems}
\author{Stefanie Russ} 
\affiliation{Institut f\"ur Theoretische Physik III, Justus-Liebig-Universit\"at
  Giessen, D-35392 Giessen, Germany}
\author{Jesper Mellenthin} 
\affiliation{Laboratoire de Physique de la Mati\`ere Condens\'ee, Ecole Polytechnique, F-91128 Palaiseau, France}
\date{\today}

\draft
\begin{abstract} 
We apply periodic-orbit theory to calculate the integrated density of states $N(k)$ from the periodic orbits of pseudointegrable polygon and barrier billiards. We show that the results agree so well with the results obtained from direct diagonalization of the Schr\"odinger equation, that about the first $100$ eigenvalues can be obtained directly from the periodic-orbit calculations in good accuracy.
\end{abstract}
\maketitle

The motion of a classical particle in a billiard system can show regular, chaotic or intermediate behavior, depending on the billiard geometry. A potential well of the same geometry as the corresponding classical billiard -- a quantum billiard -- reflects this behavior in the properties of its eigenvalues and eigenfunctions. A hallmark in the theory of chaotic systems is Gutzwiller's trace formula~\cite{gutzwiller,mehtabuch}. It expresses the density of quantum mechanical eigenstates $g(k)$ semiclassically by a weighted sum over all classical periodic orbits $i$ of lengths $\ell_i$ and thus represents an intrinsic link between the classical and the quantum mechanical properties of a given system. Since the implementation of Gutzwiller's trace formula, periodic-orbit theory has been a subject of permanent interest (for a recent review see~\cite{stoeckbuch}). However, in chaotic systems, the formula cannot be easily applied since the number $N(\ell)$ of periodic orbits with lengths smaller than a given value $\ell$ increases exponentially with  $\ell$, leading to a divergence of the trace formula. Therefore, cut-offs have to be compensated by sophisticated techniques so that the practical application of the trace formula up to now has been very limited and in most cases only reproduces the smoothed density of states and about the $20-40$ lowest individual eigenvalues~\cite{wintgen91,aurich88,steinerPA}. 
Only for the special case of the hyperbola billiard, where after a suited rearrangment of the orbits, most contributions were made to cancel, about $150$ eigenvalues have been reproduced~\cite{siebstein}.

In this paper, we concentrate on pseudointegrable billiards~\cite{pseudo1,pseudo2}, whose spectral properties as e.g. the level statistics have been found intermediate between chaotic and integrable billiards~\cite{cc,shudoshim,shudoetal,ys2003}. Systems with rough boundaries may have chaotic or pseudointegrable classical dynamics, so that the theoretical understanding of pseudointegrable systems is as important as the one of chaotic systems. One prominent application of quantum billiards is e.g. the band gap of Andreev billiards where the density of states is of major importance and where pseudointegrable billiards behave similar to chaotic billiards~\cite{andreev}. Formulas equivalent to Gutzwiller's trace formula have also been established for regular~\cite{tabor} and for pseudointegrable billiards~\cite{pseudo1}. In these billiards, the number of periodic orbits smaller than $\ell$ only increases as $N(\ell)\sim\ell^{2}$~\cite{biswas1,biswas2,biswas3}, which diminishes the divergence problems.
In this paper, we will show how the divergence problems can be overcome in these billiards and how the density of states and about the $100$ lower eigenvalues can be calculated by periodic-orbit theory.

Figure~\ref{bi:geo} shows some pseudointegrable geometries considered in this work together with some periodic orbits. Whereas rectangular systems are integrable, i.e. the motion of a particle in a rectangular billiard shows regular classical dynamics and the equations of motion can be integrated, pseudointegrable billiards are polygons with a certain number of rational angles $\varphi_i=n_i\pi/m_i$, with $n_i, m_i\in N$ and at least one $n_i>1$. Also barrier billiards~\cite{barrier} belong to this class (see Fig.~\ref{bi:geo}(d)), as a barrier can be considered as part of the boundary with an inner angle of $2\pi$. Pseudointegrable billiards are not integrable due to singularites arising at the salient corners and are classified by their genus number 
\begin{equation}\label{genus}
G = 1 + \fr{M}{2} \sum_{i=1}^{J} \fr{n_i-1}{m_i}.
\end{equation}
Here, $J$ is the number of angles and $M$ is the least common multiple of the $m_i$. In the geometries considered here, it is easy to see that every angle of value $3\pi/2$ or $2\pi$ increases $G$ by a value of $1$, whereas the angles of $\pi/2$ do not contribute.

We are interested in the density of states $g(k)$. It is well-known that in differential form $g(k)$ reads
\beq
g(k)=g_0(k)+g_{\mathrm{osc}}(k),
\eeq
where $k^{2}\equiv 2mE/\hbar^{2}$ and $m$ is the mass of the quantum mechanical particle in the potential well. $g_0(k)$ is a smooth term that can be obtained via the well-known Weyl formula~\cite{weyl} from the geometrical properties of the system. It does not require the knowledge of the individual orbits.
Hence, the calculation of $g(k)$ reduces to the oscillating part $g_{\mathrm{osc}}(k)$ that is (for the billiards considered here) connected to the lengths and the areas of the periodic orbits. In both, integrable and pseudointegrable systems, the periodic orbits form families of fixed lengths $\ell_i$, which means that the starting point of an orbit can be shifted to at least one direction along the boundary without changing its length (see Fig.~\ref{bi:geo}). Accordingly, the trajectories of all orbit families cover a finite area $A_i$ in phase space, where the index $i$ counts the different families. 
Gutzwillers trace formula has also been extended to pseudointegrable billiards~\cite{pseudo1,biswas3}. For the billiards considered here, where all orbits have an even number of reflections at the boundary walls, the boundary conditions do not play a role and the formula reads 
\beq\label{biswas}
g_{\mathrm{osc}}(k) = \sqrt{\fr{k}{2\pi^{3}}}\sum_{i} \frac{A_i}{\ell_i^{1/2}} \cos(k\ell_i-\frac{\pi}{4})
\eeq
and applies also for integrable billiards. Second-order contributions coming e.g. from diffractive orbits have been neglected and the sum is carried out over all primitive (non-repeated) orbit families $i$ and over its repetitions with multiple lenghts. Since the boundary conditions do not enter into Eq.~(\ref{biswas}), we concentrate in the following on Dirichlet boundary conditions. 

Even though the number of orbits $N(\ell)$ below a given length $\ell$ increases only quadratically with $\ell$, one can easily show that also the trace formula for pseudointegrable systems, Eq.~(\ref{biswas}), diverges and therefore could not be used so far to calculate the density of states. However, since the divergence is weak, we can use a simple trick to achieve convergence, namely by considering the fluctuations of the integrated density of states, 
\begin{eqnarray}\label{DOS}
&& N_{\mathrm{osc}}(k) = \int g_{\mathrm{osc}}(k)\,dk = \nonumber\\
&& =\frac{1}{\sqrt{2\pi^{3}}} \sum_{i} A_i
\left\lbrace\sqrt{\frac{k}{2 \ell_i^{3}}}\left[\sin(\ell_i k) - \cos(\ell_i k)\right] \right. \nonumber\\
&& +\left. \frac{\sqrt{\pi}}{2\ell_i^{2}}\left[\mathrm{FrC}\left(\sqrt{\fr{2\ell_i k}{\pi}}\,\right) - \mathrm{FrS}\left(\sqrt{\fr{2\ell_i k}{\pi}}\,\right)\right]\right\rbrace,
\end{eqnarray}	
where $\mathrm{FrS}(x)$ and $\mathrm{FrC}(x)$ are the Fresnel sine and cosine integrals, respectively, which can be evaluated numerically. Replacing the sum over $i$ by an integral over $g(\ell)\,d\ell$ and introducing the orbit density $g(\ell)=dN(\ell)/d\ell\sim\ell$, one can verify easily that the factors of $\ell_i^{-3/2}$ and $\ell_i^{-2}$ ensure the convergence of Eq.~(\ref{DOS}).

We first test the formula on the rectangular billiard where both, the orbits and the quantum mechanical eigenvalues $k^{2}_{\nu_i,\nu_j}$ are known exactly, $k^{2}_{\nu_i,\nu_j} = \pi^{2}(\nu_i^{2}/L_x^{2} + \nu_j^{2}/L_y^{2})$ with positive integers $\nu_i$ and $\nu_j$, and the side lengths $L_x$ and $L_y$ of the rectangle. The orbit lengths are $\ell_{\nu_i,\nu_j}=2 \left[(\nu_i L_x)^2+(\nu_j L_y)^2\right]^{1/2}$ and the areas are $A_i=2A$ for the neutral orbit families (the simplest orbit families that bounce between two parallel walls~\cite{sieber93}) and $4A$ for all other families, where $A$ is the geometrical area of the system.   
In Figs.~\ref{bi:rectdos}(a,b) we compare the integrated density of states $N^{PO}(k)=N_0(k)+N_{\mathrm{osc}}(k)$ (straight lines) calculated from Eq.~(\ref{DOS}) with the corresponding density of states $N^{\rm{EV}}(k)$ (circles), which has been obtained from the exact eigenvalues. $N_0(k)$ is taken from Weyl's formula (for Dirichlet boundary conditions),
\beq\label{weyl}
N_0(k) = \frac{A}{4\pi} k^{2} - \frac{\Gamma}{4\pi} k + \frac{1}{24}\sum_i(\frac{\pi}{\varphi_i}-\frac{\varphi_i}{\pi}),
\eeq
where $A$ is the area, $\Gamma$ the boundary length of the billiard and the sum runs over all corners of angles $\varphi_i$. 
One can see in Fig.~\ref{bi:rectdos} that $N^{PO}(k)$ gives the expected staircase function and the agreement to $N^{\rm{EV}}(k)$ is excellent. As shown in the upper curve of Fig.~\ref{bi:rectdos}(b), it even allows to obtain the eigenvalues directly from the steps in $N^{PO}(k)$. Each eigenvalue is positioned at one of the steps of $N^{PO}(k)$, which we verified for the first $1500$ eigenvalues (until $k^{2}=1$).

In pseudointegrable billiards, the areas are different for the different families and normally much smaller than in integrable billiards, while the number of periodic orbits is larger. In Fig.~1(b), it is demonstrated how the periodic orbits can be labeled according to their numbers of transversals of the different segments: the system has two $x$- and two $y$-segments and the orbit shown here can be labelled as $(2,6,2,0)$, where the numbers design the number of transversals of the segments $x_1$, $x_2$, $y_1$ and $y_2$ (see also Ref.~\cite{biswas3}). The orbit length and angle can be calculated from this information, whereas the area has to be calculated numerically. However, unfortunately not all combinations of integer transversals exist in the pseudointegrable systems, because many hypothetical orbits are pruned by the shielding of the corners. So, it can be seen easily in Fig.~1(b) that an orbit $(2,0,2,0)$ would not be possible in this geometry (however, this orbit exists in the geometry of Fig.~1(a)). Even though the numerical calculations can be done in high precision for up to about $50 000$ orbits in reasonable computation time~\cite{jesper2003}, we first want to investigate the stability of the results against numerical errors. To this end, we compared for the pseudointegrable billiards the orbits found in forward and backward direction. The maximum errors of the areas of the first $40000$ periodic orbits (about $300$ reflexions at the system walls including repetitions) are around $0.1\,\%$ and we found no hints that orbits might be lost. Accordingly, we have first tested the robustness of the results by disturbing the orbits of the rectangular system by errors taken from a narrow Gaussian distribution of $\sigma=0.05$ and a maximum error of $0.1 \%$ (see Fig.~\ref{bi:rectdos}(b) shifted down by a value of $5$ in the lower curve (dotted line)). One can see that on the average, $N^{PO}(k)$ still agrees very well with $N^{EV}(k)$, but that the shape of the staircase is smeared out already by these quite small errors, so that the eigenvalues can no longer be determined from the steps in $N^{PO}(k)$. The loss of about $1\%$ of the orbits (as well as their repetitions) would be less disturbing and would only lead to very slight deviations in the step function. We also checked if the number of calculated orbits is large enough and found that for all considered systems, the results are stable beyond the first $10^4$ orbits. 
 
Finally, we turn to the pseudointegrable billiards. As before, we apply Eq.~(\ref{DOS}) to obtain the density of states $N^{PO}(k)$ from periodic-orbit calculations. For comparison with the quantum mechanical eigenvalues, we solve the Schr\"odinger equation for the potential wells of Fig.~\ref{bi:geo} in its discretized version,  
\beq\label{discrete}
\sum_{i',j' (neighbors)}\left(\Psi_{n}(i',j')-\Psi_{n}(i,j)\right) = -\left( \fr{k_n}{\nu}\right)^{2}\Psi_{n}(i,j), \eeq
where the indices $(i,j)$ refer to points of a square lattice with $\nu$ lattice points per unit length. The sum over $(i',j')$ runs over all nearest neighbors of $(i,j)$. Equation~(\ref{discrete}) depends on the usual Schr\"odinger equation, $\Delta\Psi_n(i,j) = -k_n^{2}\Psi_n(i,j)$, by a second-order Taylor expansion of the left-hand side up to the quadratic term. The errors due to the discretization are the higher orders of $\nu$ and thus decrease with increasing $\nu$, while the eigenvalues are transformed via $k_n^{2}\to (k_n/\nu)^{2}$. We used a resolution of $\nu=4$, where the errors arising from the discretization (as calculated for the rectangle) are smaller than $0.03$ percent for the first $100$ eigenvalues. Equation~(\ref{discrete}) describes a matrix problem that we diagonalized by the Lanczos algorithm, yielding the numerical Lanczos eigenvalues $k^{2}_{L}$ and the corresponding density of states $N^{EV,L}(k)$. In Fig.~\ref{bi:pseudo1dos}, we compare for the polygone billiards of Figs.~\ref{bi:geo}(a-c) $N^{PO}(k)$ with $N^{EV,L}(k)$. First, in Fig.~\ref{bi:pseudo1dos}(a), we show a larger part of the energy spectrum, where we can see that both, $N^{PO}(k)$ (straight lines) and $N^{EV,L}(k)$ (symbols) agree again very well. Even though the steps in $N^{PO}(k)$ are smeared out by the numerical inaccurencies as described above, we can use them to obtain the individual eigenvalues from $N^{PO}(k)$. To this end, we fitted $N^{PO}(k)$ by a least square fit to a step function with constant integer values of the step heights. The positions $k^{2}_i$ of the steps were chosen by minimizing the quadratic deviation to $N^{PO}(k)$. The step functions obtained this way, $N_{\rm{fit}}^{PO}(k)$, are shown in Fig.~\ref{bi:pseudo1dos}(b) (solid lines). 
We can see that indeed, for about the first $100$ eigenvalues, the agreement of the Lanczos eigenvalues with this fitted step function $N_{\rm{fit}}^{PO}(k)$ is quite good. Nearly all eigenvalues are located right at the steps of $N_{\rm{fit}}^{PO}(k)$, and we can obtain at least the first $100$ eigenvalues $k^{2}_{PO}$ by periodic-orbit theory. In Tab.~\ref{tab:1}, we show as an example the 91st to the 100th eigenvalue. We calculated the number $N_m$ of mismatches among the first $100$ levels, i.e. the number of cases where the $i$th periodic-orbit eigenvalue lies closer to the $(i+1)$th or $(i-1)$th Lanczos value than to the $i$th one. $N_m$ is also given in the table and lies around $20$ percent. Naturally, the mismatches occur at energies, where the level distance is particularly small and $N_m$ is therefore highest for the system of Fig.~\ref{bi:geo}(a), where the density of states increases fastest and the level distances are thus smallest.

As last example, we consider the barrier billiard of Fig.~\ref{bi:geo}(d) for three different heights of the barrier, $h=10$, $h=50$ and $h=100$. In this case, the discretization of the lattice is a cruder approximation than before, since the barriers which should be of thickness zero, always occupy one grid point. In Fig.~\ref{bi:pseudo2dos}, we can see that at least for the systems with barrier heights $h=10$ and $h=50$, the agreement between the periodic-orbit and the Lanczos results is again very good with mismatches even below $20$ percent (see Tab.~\ref{tab:2}). Only for the billiard with the largest barrier height, the mismatches are larger which is, however, most probably due to the discretization procedure and not to the periodic-orbit calculations. A comparison to the eigenvalues calculated by some other procedures will be interesting.

Finally, we want to compare our results to some other methods that were used in the past to determine eigenvalues of systems with chaotic classical dynamics by periodic-orbit theory. The first methed used e.g. in~\cite{aurich88} is a ''Gaussian smearing'' of $g(k)$. This means that convergence of the trace formula can be achieved by multiplying each term of Eq.~(\ref{biswas}) with the additional factor of $\exp(-l_i^2\epsilon^{2}/2)$ (where $\epsilon$ must be small). As a consequence, the delta peaks of $g(k)$ are transformed into Gaussian functions of shapes $\sim\exp(-k^2\epsilon^{-2}/2)$. We tested this method also on our systems and found that it works very well for the integrable rectangular systems, but not for the pseudointegrable systems. Similar as in our method, the small errors in the orbit areas lead to a large noise that is still increased by the ''Gaussian smearing''. Contrary to our method, where we were able to apply a clearly defined fit procedure to eliminate the noise from the steps in $N(k)$, it is not possible to find the Gaussian functions in $g(k)$ by a simple rule. Another method used in~\cite{siebstein} calculates the eigenvalues from the zeros of the so-called dynamical zeta function that contains all orbit information. In very special cases, where every large primitive orbit can be decomposed into series of few small orbits, the dynamical zeta function can be calculated very easily. However, this condition is only fulfilled in rare cases and demands as minimal conditions that the orbits can be labelled into fundamental building blocks, where each combination of the blocks exists. We have seen that in the case of pseudointegrable billiards, the periodic orbits appear in a very unsystematic way and many hypothetical orbits are pruned, so that we think that this method is not helpful in our case. 

In summary, we have shown how the convergence problems of the trace formula can be overcome in systems, where the number of periodic orbits below a given length $\ell$ increases at most quadratically with $\ell$, e.g. for integrable and for pseudointegrable billiards. We have shown that the integrated density of states can be reproduced in very good accuracy for several hundred eigenvalues. The calculations are very sensible to numerical errors, so that already error bars of about $0.1\%$ destroy the shape of the step function of $N^{PO}(k)$ such that the steps are smeared out. In integrable billiards, where the orbits are known exactly, the eigenvalues can be found directly from the steps in $N^{PO}(k)$ (which we tested for the first $1500$ eigenvalues). In pseudointegrable billiards, even if the curves follow all fluctuations of the spectra very well, a fit technique has to be used in order to find the first $100$ individual eigenvalues. The results are very promising and show that the quantum mechanical density of states can indeed be gained, using as input solely the classical periodic orbits of the pseudointegrable billiards. It will be very interesting to apply this procedure to real mesoscopic structures, where the change of the density of states as response to a change in geometry is of great importance.

We would like to thank A. Bunde for a careful reading of the manuscript and valuable remarks.

\newpage
\unitlength 1.85mm
\vspace*{0mm}
\begin{figure}
\begin{picture}(45,55)

\put(14,42){\makebox(1,1){\bf\Large (a)}} 
\put(35,42){\makebox(1,1){\bf\Large (b)}} 
\put(16,14){\makebox(1,1){\bf\Large (c)}} 
\put(35,14){\makebox(1,1){\bf\Large (d)}} 

\put(-1,25){\makebox{\includegraphics[width=3.3cm, height=5.08cm]{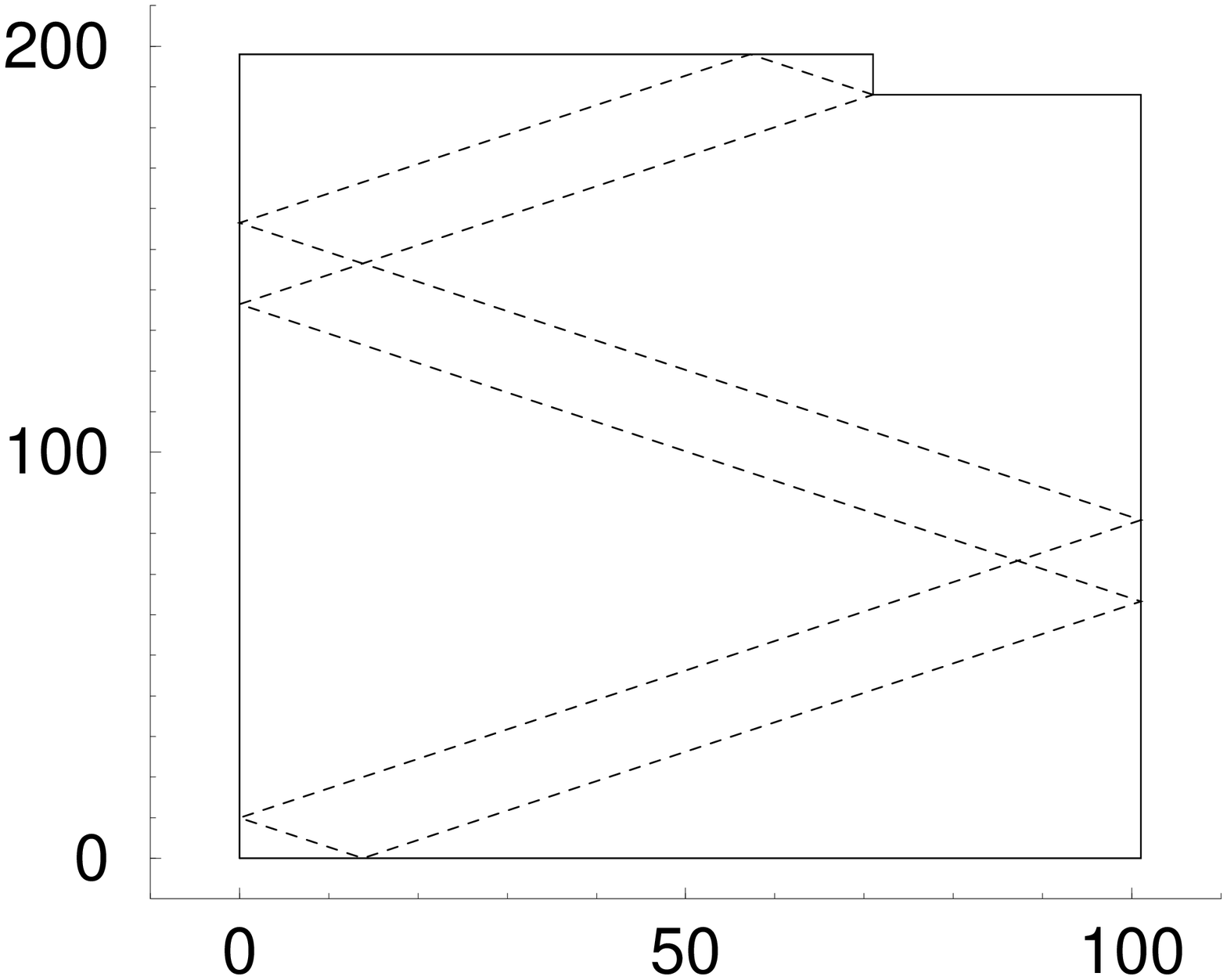}}}
\put(23,25){\makebox{\includegraphics[width=3.3cm, height=5.08cm]{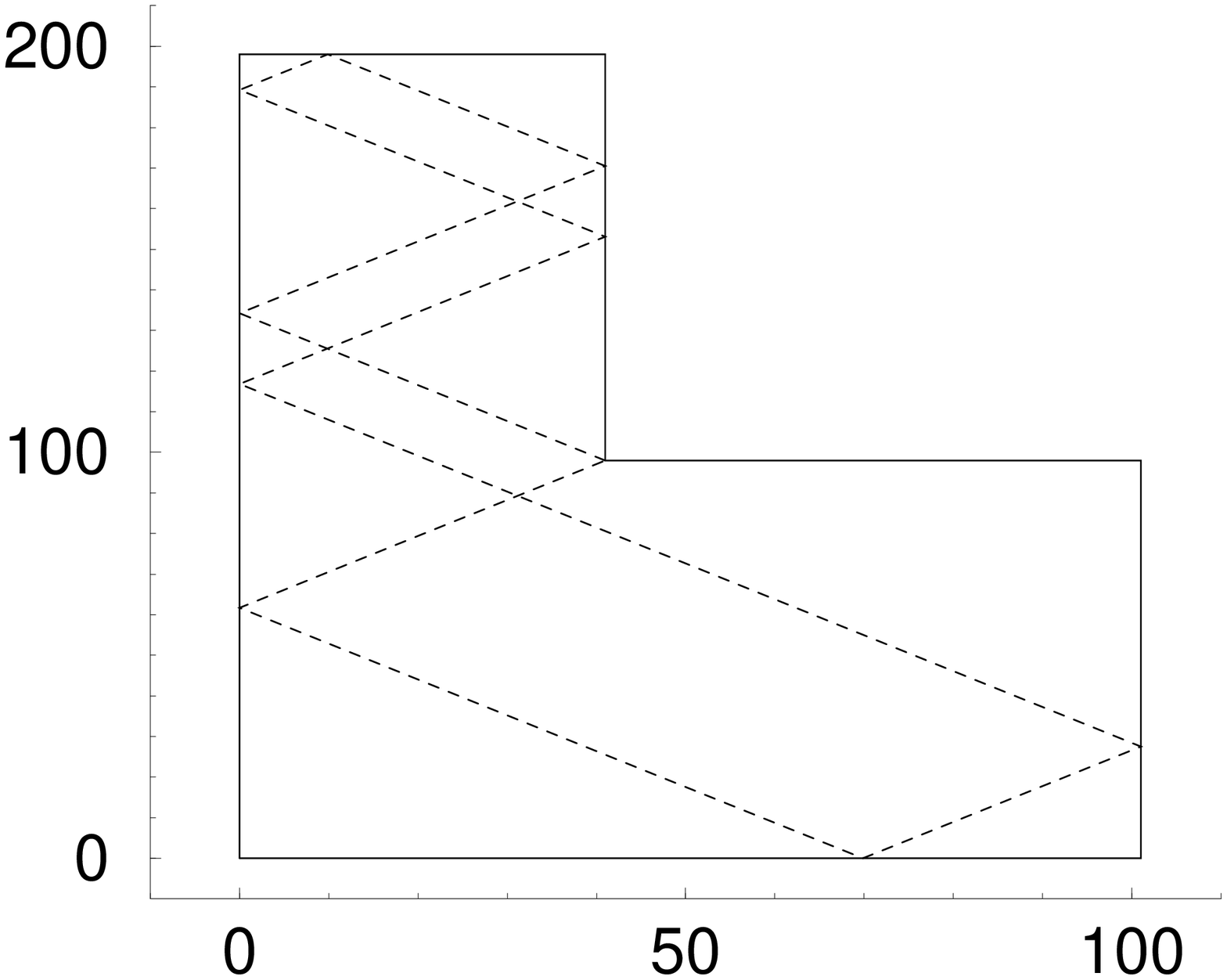}}}
\put(-1,0){\makebox{\includegraphics[width=3.3cm, height=5.08cm]{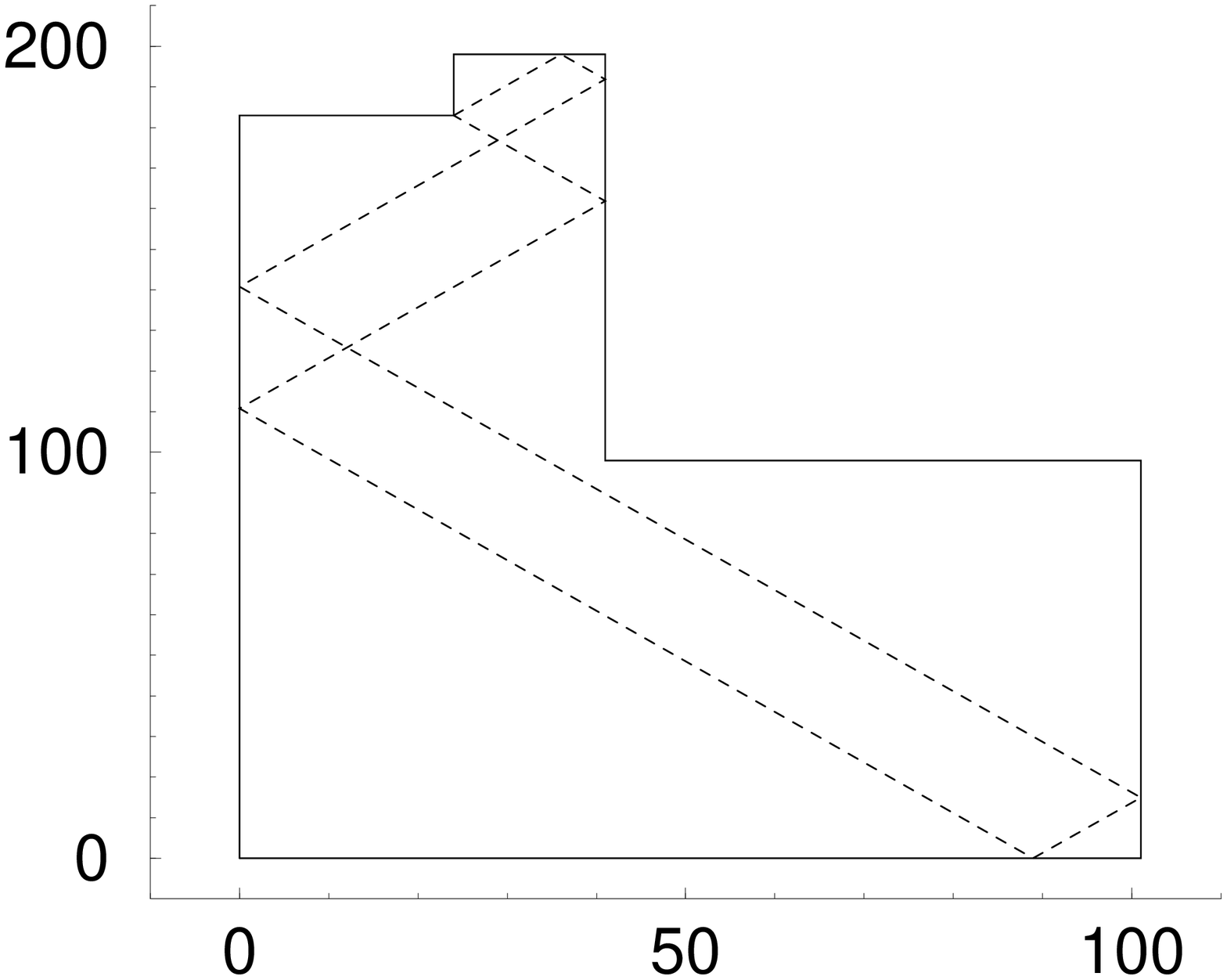}}}
\put(23,0){\makebox{\includegraphics[width=3.3cm, height=5.08cm]{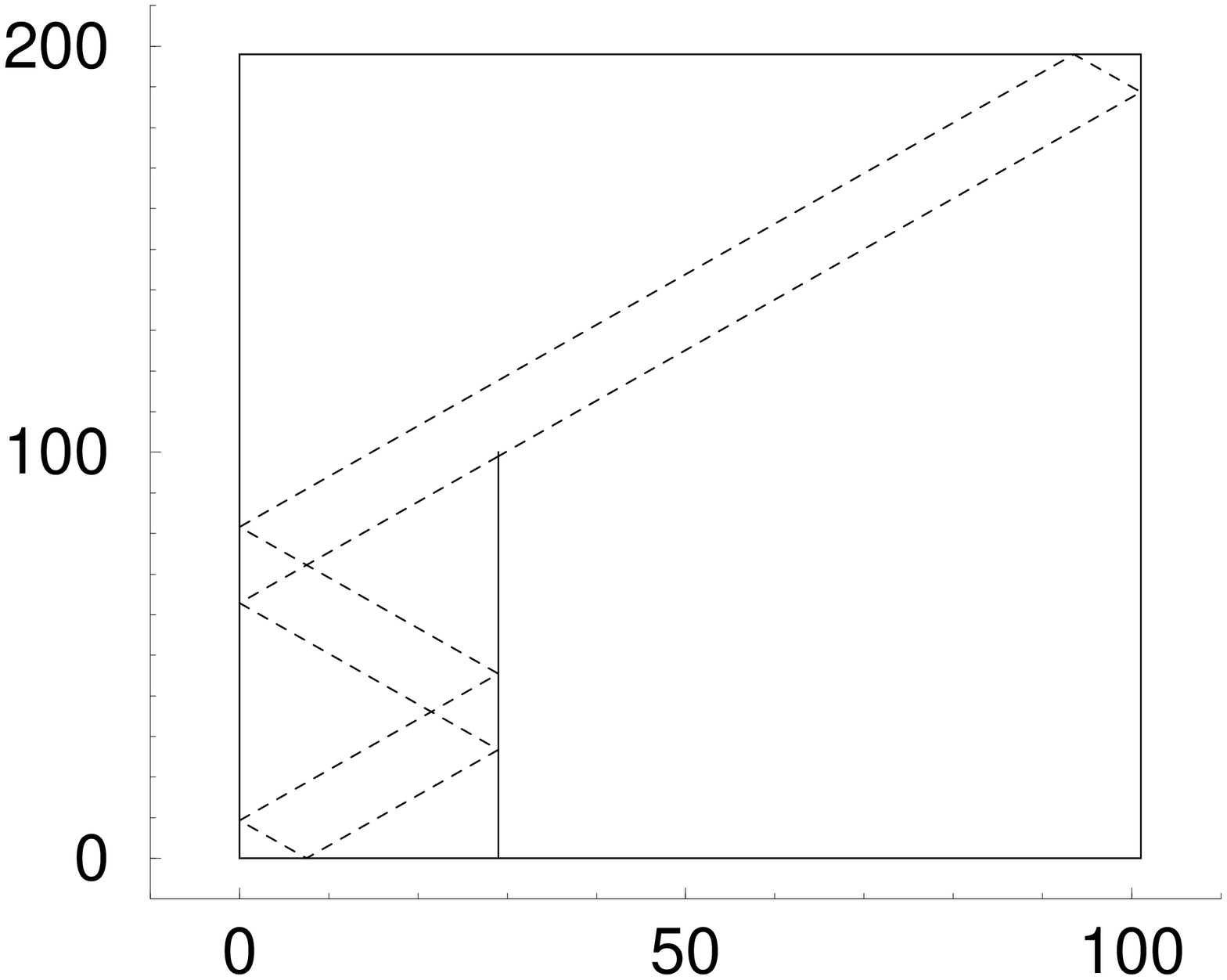}}}

\put(25,50){\line(1,0){17}}      
\put(42,49.5){\line(0,1){1}}      
\put(25,49){\line(0,1){2}}      
\put(32,49.5){\line(0,1){1}}      
\put(43,29){\line(0,1){18}}      
\put(42,29){\line(1,0){2}}      
\put(42.5,47){\line(1,0){1}}      
\put(42.5,37.5){\line(1,0){1}}      
\put(42,51){\makebox(1,1){\bf $x_1$}} 
\put(32,51){\makebox(1,1){\bf $x_2$}} 
\put(44.5,37.){\makebox(1,1){\bf $y_2$}} 
\put(44.5,46.5){\makebox(1,1){\bf $y_1$}} 

\end{picture}
\caption[]{\small Shapes of the pseudointegrable billiards considered in this work. (a,b) $L$-shaped billiards (polygon billiards of genus number $G=2$), (c) a polygon billiard of $G=3$ and (d) the barrier billiard ($G=2$), which is calculated for different heights $h$ of the barrier. One periodic orbit for each geometry is also shown (dashed lines). In (b) the different segments are shown that can be transversed by the periodic orbits. 
}
\label{bi:geo}
\end{figure}

\begin{figure}
\begin{picture}(40,40)
\put(5,33){\makebox(1,1){\bf\Large (a)}} 
\put(25,33){\makebox(1,1){\bf\Large (b)}} 
\put(1,0){\makebox{\includegraphics[width=7cm]{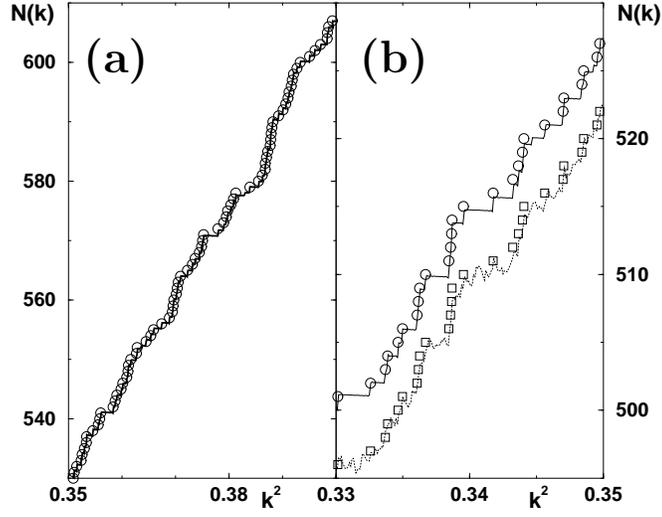}}}
\end{picture}
\caption[]{\small (a,b) Density of states $N^{PO}(k)$ calculated from the periodic orbits (solid lines) and $N^{EV}(k)$ calculated  from the exact eigenvalues (circles) of a rectangular system of side lengths $L_x=101$ and $L_y=198$. In (b), the upper curve compares the steps in $N^{PO}(k)$ to the eigenvalues. In the lower curves (shifted down by a value of $5$), the areas of the orbits are disturbed by errors of up to $0.1\%$ and it can be seen that the steps in $N^{PO}(k)$ (dotted line) are strongly disturbed already by these small error bars. 
}
\label{bi:rectdos}
\end{figure}

\begin{figure}
\begin{picture}(40,40)
\put(5,33){\makebox(1,1){\bf\Large (a)}} 
\put(25,33){\makebox(1,1){\bf\Large (b)}} 
\put(1,0){\makebox{\includegraphics[width=7cm]{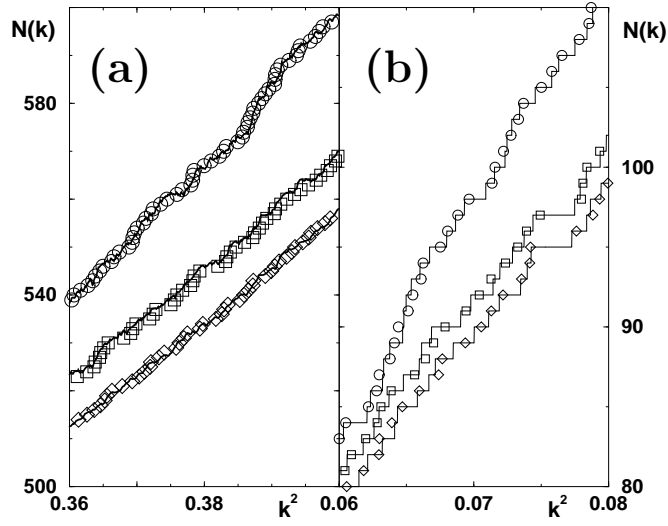}}}
\end{picture}
\caption[]{\small (a,b) Density of states $N^{PO}(k)$ calculated from the periodic orbits (solid lines) and $N^{EV,L}(k)$ from the Lanczos eigenvalues for the  $L$-shaped systems of Fig.~\ref{bi:geo}(a) (circles) and Fig.~\ref{bi:geo}(b) (squares) and the system of genus number $G=3$ of Fig.~\ref{bi:geo}(c) (diamonds). In (b), the fitted step function of $N_{\rm{fit}}^{PO}(k)$ is compared to the Lanczos eigenvalues. For a better overview the data of the geometries of Figs.~\ref{bi:geo}(b,c) have been shifted upwards by values of $150$ in (a) and by $25$ in (b).
}
\label{bi:pseudo1dos}
\end{figure}

\begin{figure}
\begin{picture}(40,40)
\put(5,33){\makebox(1,1){\bf\Large (a)}} 
\put(25,33){\makebox(1,1){\bf\Large (b)}} 
\put(1,0){\makebox{\includegraphics[width=7cm]{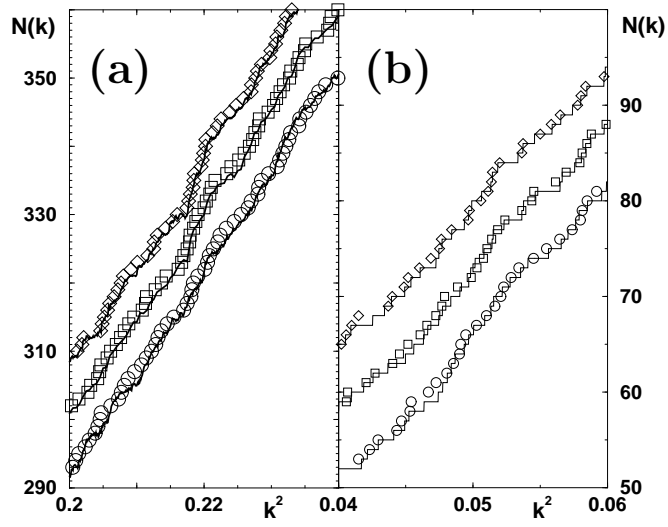}}}
\end{picture}
\caption[]{\small (a,b) Density of states $N^{PO}(k)$ calculated from the periodic orbits (solid lines) and $N^{EV,L}(k)$ from the Lanczos eigenvalues for the barrier billiards with barrier heights $h=10$ (diamonds), $h=50$ (squares) and $h=100$ (circles). In (b), the fitted step function of $N_{\rm{fit}}^{PO}(k)$ is compared to the Lanczos eigenvalues. For a better overview the data in both, (a) and (b) have been shifted upwards by values of $5$ ($h=50$) and by $10$ ($h=10$).}
\label{bi:pseudo2dos}
\end{figure}

\begin{table}
\begin{tabular}{|c|c|c||c|c||c|c|} \hline
   {\rule[-2mm]{0mm}{6mm}}   & \multicolumn{2}{c||}{Fig.~\ref{bi:geo}(a)} & \multicolumn{2}{c||}{Fig.~\ref{bi:geo}(b)} & \multicolumn{2}{c|}{Fig.~\ref{bi:geo}(c)} \\ \hline 
      {\rule[-2mm]{0mm}{6mm}} n & $k^{2}_{PO}$ & $k^{2}_L$ & $k^{2}_{PO}$ & $k^{2}_L$ & $k^{2}_{PO}$ & $k^{2}_L$\\  \hline 
91 & 0.0649 & 0.0651 & 0.0948 & 0.0949 & 0.0975 & 0.0978\\
92 & 0.0650 & 0.0655 & 0.0959 & 0.0960 & 0.0979 & 0.0984\\
93 & 0.0658 & 0.0661 & 0.0971 & 0.0968 & 0.0981 & 0.0990\\
94 & 0.0664 & 0.0662 & 0.0977 & 0.0977 & 0.0990 & 0.0996\\
95 & 0.0668 & 0.0675 & 0.0979 & 0.0982 & 0.0999 & 0.0998\\
96 & 0.0680 & 0.0680 & 0.0985 & 0.0988 & 0.1008 & 0.1007\\
97 & 0.0689 & 0.0688 & 0.0997 & 0.0999 & 0.1017 & 0.1019\\
98 & 0.0701 & 0.0697 & 0.1001 & 0.1004 & 0.1022 & 0.1024\\
99 & 0.0713 & 0.0713 & 0.1008 & 0.1010 & 0.1038 & 0.1045\\
100 &0.0716 & 0.0716 & 0.1013 & 0.1015 & 0.1042 & 0.1047\\
\hline
   {\rule[-2mm]{0mm}{6mm}} $N_m$  & \multicolumn{2}{c||}{24} & \multicolumn{2}{c||}{15} & \multicolumn{2}{c|}{21} \\ \hline 
\end{tabular}
\caption[]{\small Table of the 91st to the 100th eigenvalue $k^{2}_{PO}$ calculated by the trace formula compared to $k^{2}_L$ calculated by the Lanczos algorithm for the polygon pseudointegrable systems. $N_m$ is the number of mismatches during the first $100$ values, i.e. the number of cases where the $i$th periodic-orbit eigenvalue lies closer to the $(i+1)$th or $(i-1)$th Lanczos value than to the $i$th one.
\label{tab:1}}
\end{table}   

\begin{table}
\begin{tabular}{|c|c|c||c|c||c|c|} \hline
{\rule[-2mm]{0mm}{6mm}}   & \multicolumn{2}{c||}{Fig.~\ref{bi:geo}(d), $h=10$} & \multicolumn{2}{c||}{Fig.~\ref{bi:geo}(d), $h=50$} & \multicolumn{2}{c|}{Fig.~\ref{bi:geo}(d), $h=100$} \\ \hline 
      {\rule[-2mm]{0mm}{6mm}} n & $k^{2}_{PO}$ & $k^{2}_L$ & $k^{2}_{PO}$ & $k^{2}_L$ & $k^{2}_{PO}$ & $k^{2}_L$\\  \hline 
\hline
91 & 0.0647 & 0.0649 & 0.0649 & 0.0654 & 0.0662 & 0.0663\\
92 & 0.0649 & 0.0653 & 0.0656 & 0.0656 & 0.0674 & 0.0669\\
93 & 0.0654 & 0.0654 & 0.0658 & 0.0660 & 0.0678 & 0.0683\\
94 & 0.0655 & 0.0657 & 0.0675 & 0.0672 & 0.0688 & 0.0693\\
95 & 0.0663 & 0.0661 & 0.0678 & 0.0679 & 0.0692 & 0.0697\\
96 & 0.0672 & 0.0673 & 0.0680 & 0.0683 & 0.0698 & 0.0704\\
97 & 0.0678 & 0.0678 & 0.0692 & 0.0696 & 0.0708 & 0.0709\\
98 & 0.0682 & 0.0683 & 0.0698 & 0.0698 & 0.0716 & 0.0712\\
99 & 0.0693 & 0.0692 & 0.0706 & 0.0713 & 0.0723 & 0.0727\\
100 &0.0708 & 0.0710 & 0.0722 & 0.0724 & 0.0726 & 0.0730\\
\hline
\hline
   {\rule[-2mm]{0mm}{6mm}} $N_m$  & \multicolumn{2}{c||}{16} & \multicolumn{2}{c||}{18} & \multicolumn{2}{c|}{29} \\ \hline 
\end{tabular}
\caption[]{\small Table of the 91st to the 100th eigenvalue $k^{2}_{PO}$ calculated by the trace formula compared to $k^{2}_L$ calculated by the Lanczos algorithm for the pseudointegrable barrier systems. $N_m$ is the number of mismatches during the first $100$ values.
\label{tab:2}}
\end{table}   

\end{document}